\documentclass[letter]{article}
\usepackage{epsfig}

\begin{document}

\title{ Differentiated End-to-End Internet Services using a
	       Weighted Proportional Fair Sharing TCP}

\author {Jon Crowcroft and Philippe Oechslin \\
	University College of London \\
	\{jon,p.oechslin\}@cs.ucl.ac.uk
	}

\maketitle

\begin{abstract}

In this document we study the application of weighted proportional
fairness to data flows in the Internet.  We let the users set the
weights of their connections in order to maximise the utility they get
from the network. When combined with a pricing scheme where
connections are billed by weight and time, such a system is known to
maximise the total utility of the network. Our study case is a
national Web cache server connected to long distance links. We propose
two ways of weighting TCP connections by manipulating some parameters
of the protocol and present results from simulations and prototypes. We
finally discuss how proportional fairness could be used to implement an
Internet with differentiated services.

\end{abstract}

\section{Introduction}
\thispagestyle{empty}

\subsection{Fairness}
Fairness is among the most important properties of data flows in the
Internet.  Fairness implies that whenever there is congestion at a
bottleneck, each flow going through that bottleneck gets a fair share
of the available bandwidth. TCP flows, which make up most of the
Internet's data flows, achieve at least approximate fairness by 
using congestion control
mechanisms \cite{jac88} which adapt each TCP's throughput as a  function of the
congestion.

The most common form of fairness is max-min fairness. In a max-min
fair system all connections get the same share of a bottleneck. If a
connection can not use all of its share, e.g. because it has a slower
rate in an other bottleneck, then the excess capacity is shared fairly
among the other connections. In other words, a source that is not able
to use more than one $N$th of the bottleneck's bandwidth will always
be able to send at its maximum rate.

Another form of fairness is proportional fairness. A system is
proportionally fair if any change in the distribution of the rates
would result in the sum of the proportional changes being negative. If
a source is not able to use one $N$th of the bottleneck it may still be
allocated less than its maximum, say 5\% less, if this allows a
larger, say more than 5\%, increase of the rate of another connection. 

For the exact definition of min-max and proportional fairness see
Annex A. For the rest of this paper we will be looking at weighted
proportional fairness, where each connection is associated with a
price. In that case it is not the rates that are proportionally fair
but the amount paid per rate. Thus one connection with the price of
two would get the same rate as two connections with a price of one.

Exciting results concerning weighted proportional fairness have been
published recently \cite{kel97}. One of the results is that rate
control based on additive increase and multiplicative decrease, as in
TCP, achieves proportional fairness. The other result is that in a
weighted proportionally fair system where the weights are the prices
the users pay per time unit, when each user chooses the price that
maximises the utility he or she gets from the network, the system
evolves to a state where the total utility of the network is
maximised. It is a typical example of local optimisations leading to a
global optimum. This property even holds when the exact function
relating utility to the bandwidth received by a user is unknown and
different for each user. The only constraint on that function is that
the utility has to be an increasing, concave and differentiable
function of the bandwidth, which happens to be one of the definitions
of elastic traffic \cite{she95}.

\subsection{The differentiated services Internet and weighted
proportional fairness}
The above is a very interesting result in the context of service
differentiation in the Internet. Indeed it has been recognised that
the needs of the users of the Internet are not all the same and that
the current solution which provides the same service to all users is
not optimal. Several solutions have been proposed for an Internet with
differentiated services \cite{cla97,nic97}. Many solutions
aim at providing a small number of service classes (typically two or
three) with well defined quality and prices. This is implemented by
using multiple queues (one per service class) in most gateways of
the network. The price being paid for a given service (e.g. premium
service) does not depend on the congestion of the network. The network
provider thus has to over-provision its premium service to make sure
that there is always enough capacity available. This is usually done by
selling only a small fraction of the network capacity for premium
service use. The network provider also has to implement some connection
acceptance control algorithm (CAC) to makes sure that there are never
too many users using the premium service at the same time.

Weighted proportional fairness, where the weight of each flow is given by the
price being paid, would allow Internet Service Providers (ISPs) to
implement an Internet with differentiated services without the
limitations of the solutions cited above. There would be no need for
multiple queues in the network or for CAC schemes at the border of the
network. The quality perceived per price payed would vary in function
of the congestion of the network, thus always allowing a maximised
utilisation of the network.

\subsection{Related Work}

In the section above, we have cited two proposals for the
implementation of differentiated services in the Internet. This is a
relatively new field and there other proposals being discussed in the
Int-Serv working group of the Internet Engineering Task Force. One of
the proposals \cite{wan97} is similar to our's in the sense that it
also aims at providing a fair share of the network capacity to
the users. Concerning the behaviour of TCP connections, the work in
\cite{bal97} is somehow related to our work. The authors propose to
integrate a group of TCP connections between two end points for better
efficiency of HTTP transfers. The integrated group then behaves like a
single TCP connection with regard to congestion control. This is
orthogonal to our approach of having one connection behaving like $N$
connections to achieve a better quality of service, and one could
imagine combining both approaches.

\subsection{Goal and Overview}

It is the goal of this paper to explore the practical implications of
implementing weighted proportional fairness for our specific study case and for
the Internet in general. The rest of this document is organised as
follows. Section \ref{sec:uk_cache} describes the study case. Section
\ref{sec:rec_buf} describes how proportional fairness can be achieved
by adjusting TCP receive buffers in a WWW cache server.\footnote{Since
this work was done, the authors were made aware of a similar approach,
published in\cite{kal98}.} The next
section presents results for a more general solution consisting in
modifying the congestion control algorithm in TCP. Section
\ref{sec:billing} addresses the problems of billing and policing and
makes a proposition of how to introduce weighted proportional fairness
in todays Internet. Finally, the last
section concludes the paper by discussing the various results in the
light of an overall emerging pricing architecture for differentiated
services.

\section{The UK WWW cache system}
\label{sec:uk_cache}

The context in which we want to study proportional fairness is the UK
World Wide Web caching system. Caches are used in the Internet to
accelerate the access to Web pages by storing them in cache servers
located close to the end users. This is particularly efficient for
pages which would have to be fetched over transoceanic links which are
congested most of the time. The UK cache system is made of a few
coordinating cache servers scattered throughout the UK. The root
server uses a dedicated transatlantic link to fetch data from the
US. That dedicated link is usually less congested than the link used
for general traffic. This acts as an incentive for people to use the
cache which in turn augments its efficiency and thus reduces the total
amount of data fetched over the ocean.

\begin{figure}
\begin{center}
\epsfig{file=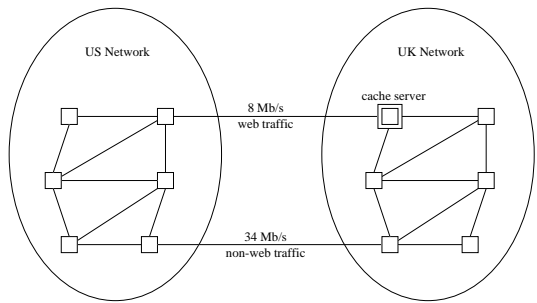,width=7cm}
\caption{Study case configuration for the UK WWW cache service}
\end{center}
\label{fig:uk_cache}
\end{figure}

If end-to-end weighted proportional fairness is to be provided to the users of
the cache service, then obviously the cache server would be an optimal
place to implement it. The bottleneck which has to be shared is the
transatlantic link to the US. On this link data is flowing towards the
cache servers which are thus at the receiving end of the TCP
connections.


We do not consider the case where the document is already
stored in the cache server as the national network is considered to be
uncongestioned. 

In the next two sections we will explore two ways of providing
weighted proportional fairness. One way consists in modifying the receive
buffer of the TCP connections to limit their throughput. The other
solution is to modify the aggressiveness of the connection control
algorithm in TCP.


\section{Limiting the Receive Buffer}
\label{sec:rec_buf}

The first method we explore hinges on limiting the size of the receive
buffers on the main cache server for connections from original servers
to the cache server. In Section \ref{sec:multcp} we will see a second
method based on modifications of TCP's congestion control.

\subsection{Description}

The receive buffer of a TCP socket limits the maximum window that can
be advertised by the receiver. As there can never be more than one
window worth of data in flight between the sender and the receiver,
the receive buffer size limits the throughput $T$ of a TCP connection to
\[T \leq \frac{B_R}{{\rm R}}\]
where $B_R$ is the receive buffer size and $R$ is the round trip
time. 

Limiting the receive buffers of a set of connections terminating at
one host provides proportional fairness if all connections
share a common bottleneck. Fortunately, in our study case, the bottleneck
is the transoceanic link and the Web cache is sitting at the receiving
end. In that particular case the solution has the
advantage that it only requires modifications on the cache servers
and none on the other endpoints of the connections.

\begin{figure}
\begin{center}
\epsfig{file=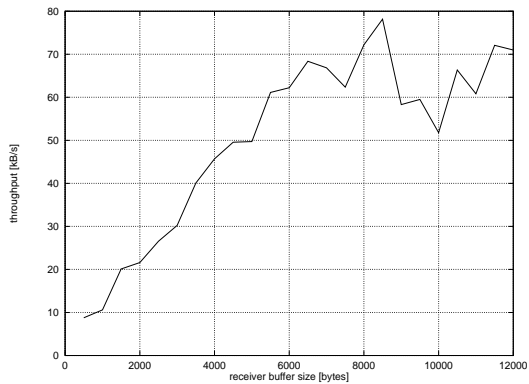,width=7cm,height=5cm}
\caption{Throughput as a function of the receive buffer size }
\end{center}
\label{fig:recw}
\end{figure}

The difficulty with limiting the receive buffer is that it sets an
absolute maximum to the throughput of a connection. Proportional
fairness, however, requires that the bandwidth be set in proportion to
a fair share of the available capacity. As the number of connections
changes, the fair share changes too. The consequence is that the
receive buffer sizes have to be adjusted every time a connection
starts or stops. The sum of all receive buffers should be equal to the
maximum amount of data that can be in transit. This is equal to the
bandwidth of the bottleneck multiplied by the mean RTT of all
connections. Call this amount $B$. Call the price that the user of
each connection wants to pay $k_i$. To achieve proportional fairness
each connection should be assigned a buffer of size
\[ b_i = B \frac{k_i}{\sum{k_j}} \]
All buffer sizes have to be adjusted whenever a connection starts,
stops or when a user decides to change the price she is paying.
This may incur a lot of overhead in a busy cache server. We conjecture 
that on a large server the total number of connection and the average 
price being paid will not vary rapidly,thus receive buffers may not 
have to be adjusted frequently.



Some negotiation mechanism is necessary for the users to indicate to
the cache system how much they would like to pay. This will not be
necessary every time a document is downloaded. More likely, the price
will only be adjusted when the utility perceived by the user changes,
for example when the user stops using the web for important work and
starts surfing random places.


\subsection{Experimental Results}

We have implemented a prototype of a cache server with variable
receive buffers. Users can select the size of the buffer through a
fill-out form on the server. Figure \ref{fig:recw} shows the
throughput obtained when transferring the same amount of data over a
long distance link for various buffer sizes. In this experiment we see
that the throughput increases linearly with the receive buffer size up
to a size of 6kB. At that point the throughput starts to be limited by
the losses. The variation of the load in the network explains the
evolution of the plot for values above 6kB. It also explains the small
non-linearities in the lower part of the plot. The variation of the
load in the network causes variations in queue sizes which in turn
affect the throughput by varying the round trip time.  Note also that
TCP receive windows can only be closed at the rate at which packets
are received, which limits the rate of adaption of this approach.

The solution provided in this Section works well when all connections
share a same bottleneck. For a more general case of a network with 
multiple bottlenecks we look into a distributed solution to the 
problem:

\section{MulTCP, a schizophrenic TCP}
\label{sec:multcp}

This is our second method for implementing weighted proportional
fairness. It is more general as it is not limited to one specific
service, web caching and requires modifications on the end systems
only. 

\subsection{How does it work?}

MulTCP is a TCP that behaves as if it was a collection of
multiple virtual TCPs. To prevent the network from collapsing when
congestion occurs, TCP has been provided with mechanisms that will
reduce its throughput when losses are detected \cite{jac88}. From
\cite{mat97,flo91} we know that the throughput of a single TCP
connection is inversely proportional to both the square root of its
loss rate $p$ and to its round trip time $R$: \[T = \frac{C}{{\rm R}
\sqrt{p}}\]

where the exact value of $C$ depends on the approximations made. When
multiple TCP streams go through a congested gateway, they
experience approximately the same loss rate and thus get about the same
fair share of the gateway's bandwidth. An equal loss rate can be
enforced by advanced queue management techniques like RED. The share
of bandwidth given to each connection is then only biased by the
round trip times\footnote{Note that this bias towards connections with
small RTTs actually encourages the use of cache servers. Indeed, even
if the cache server has to fetch the document this results
in two connections with smaller RTTs than one direct connection with a
large RTT.}


Our goal is to design a TCP control algorithm which takes a factor
$N$ as parameter and results in a TCP connection getting the same
share of congested gateways bandwidth as $N$ standard TCPs would
get. 

A TCP goes through different phases when it starts up, experiences
loss or gets into some sort of steady state. In any of these phases,
our MulTCP has to behave like $N$ concurrent TCP connections would:

\paragraph{Slow start:} During slow start a TCP opens its congestion window
exponentially by sending two packets for every acknowledgement
received. Interestingly, $N$ TCPs doing slow start still send only two
packets per acknowledgement received. However, $N$ TCPs would start by sending $N$
single packets, resulting in $N$ acknowledgements being received and $2N$ packets
being sent out after one RTT. The same behaviour could be achieved by
MulTCP if it sent out $N$ packets at startup and then two packets for
every acknowledgement received. This, however, leads to very bursty patterns if
$N$ is large. Burst may result in bursts of losses which in turn
prevent the connection of rapidly reaching steady state. MulTCP thus
uses a smoother option. It starts like a normal TCP by sending a single
packet. After that, it sends three packets for each acknowledgement received until
it has opened its congestion window as far as $N$ TCPs would have.

After $k$ round trip times $N$ TCPs have a congestion window of $N
2^{k}$. One MulTCP sending three packets for each acknowledgement would have a
window of $3^{k}$. Thus they have the same window after $k_N$ round
trip times where \[ k_N = \frac{\log N}{\log3 - \log2}\] which happens
when the window has a size of 
\[w_N = 3^{k_N}\]

The resulting pseudo code looks like this\footnote{In an optimised
implementation the expression containing a power operation and two
logarithms could be cached or looked up in a table. Also, if we chose a
burstier approach consisting in sending for packets per
acknowledgement, the expression would simplify to $N^2$} :

\begin{verbatim}
  if (cwnd < ssthresh) { /* slow-start  */
    if (cwnd <= pow(3.0,log(N)/(log(3)-log(2))))
      cwnd += 2;
    else
      cwnd += 1;
  }
\end{verbatim}

\paragraph{Linear increase:} When the congestion window reaches {\em
ssthresh} a TCP increases its window by one packet per RTT or by
$\frac{1}{cwnd}$ per packet. $N$ TCPs increase their window by $N$
packets per RTT or $\frac{N}{cwnd}$ per packet.

\paragraph{Multiplicative decrease:} When a TCP notices congestion through
the loss of a packet it halves its congestion window, sets {\em
ssthresh} to the new value of the congestion window and goes back to
linear increase.  When $N$ TCPs are sending data and one packet is lost,
only one TCP will halve its window. Thus MulTCP, when it experiences
loss, only halves one Nth of its congestion window by setting {\em
cwnd} and {\em ssthresh} to $\frac{N-0.5}{N}$ of {\em cwnd} This
assumes that at the time of loss all $N$ virtual TCPs had the same
values for these variables. This is macroscopically true since the
fairness properties also hold between the virtual TCPs. Moreover,
looking at this in more detail, we can easily see that $N$ TCPs
experiencing a total of $k$ losses randomly distributed amongst them
end up with a sum of congestion windows which has a statistical mean
of $\left(\frac{N-0.5}{N}\right)^k$. This is equal to the congestion
window of a single TCP which reduces its window by $\frac{N-0.5}{N}$
for each loss.


\begin{verbatim}
  if (cwnd < ssthresh)
    cwnd = cwnd/2;
  else
    cwnd = cwnd*(N-0.5)/N;
  ssthresh = int(cwnd);
\end{verbatim}

Note that when the connection is in slow start it is probing the
network by doubling the window every RTT. A loss during that phase
means that the window is up to two times too large. Not reducing it by
two may result in many consecutive losses which in turn may result in
a timeout.

\paragraph{Timeout:} Timeouts occur when there are too many losses
within one RTT, such that not enough acknowledgements are received to keep the
sender sending. The protocol stalls, a timeout occurs and transmission
restarts with a slow start after the last acknowledged packet. $N$
TCPs are less prone to timeout that one MulTCP. Since the losses are
distributed over $N$ connections the probability that one TCP
experiences enough losses within one RTT to make it stall is
smaller. Moreover, if one TCP should stall, the $N-1$ others can still
go on sending. There is not much we can do here to make MulTCP like
$N$ TCPs. The fact that it has only one control loop through one
sender and one receiver makes it more vulnerable to bursts of losses
than $N$ TCPs having $N$ control loops.

The only thing we can do to reflect this is to reduce the slow-start
threshold to $\frac{N-0.5}{N}$ of its value rather than halve it. Thus
after the slow-start is over the MulTCP will have the same window as
$N$ TCPs would after one of them has done a slow-start.

\begin{figure}[htb]
\begin{center}
\epsfig{file=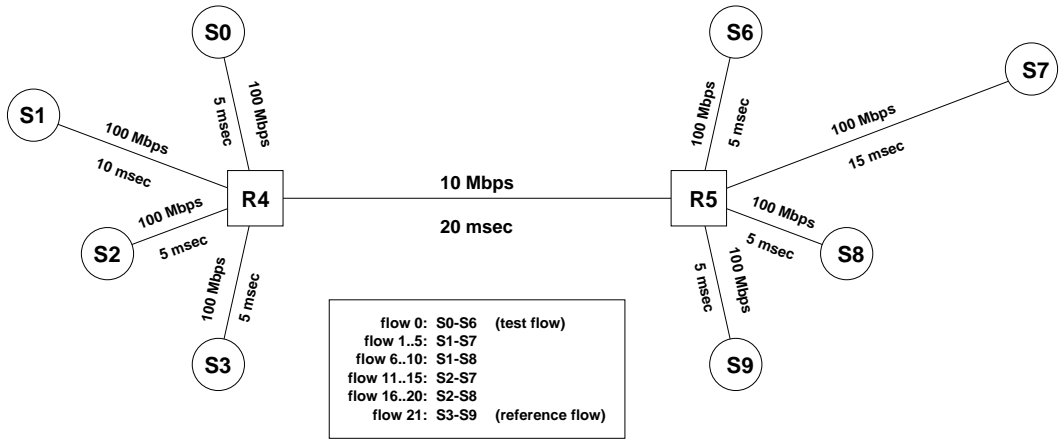,width=14cm}
\caption{The network used for simulations. The routers use RED with
thresh = 5, maxthresh = 15 and limit = 20 }
\end{center}
\label{fig:network}
\end{figure}


\subsection{Hyperinflation and congestion collapse}

There is a justified fear that if all users starting paying more for
their connections the throughput obtained for a single fair share will
become very small. Since a fair share corresponds to a normal TCP
connection and throughput is inversely proportional to the loss rate
this means that the loss rate will be high. If the value of a single 
fair share becomes small, users may want to buy many shares and thus
increase loss and drive the network into congestion collapse. 

This scenario can only happen if the price for a single fair share is set too
low. Indeed there is a finite amount of money that is spent on the 
connexions and the average loss rate can be regulated by setting the 
appropriate price for a fair share. 

\subsection{Simulation Results}

For the steady state, the above modifications lead to a theoretical
throughput which is approximatively $N$ times larger than the
throughput of one TCP for the same error rate. The development of this
result is given in the Annex B.

\begin{equation}
T =  \frac{\sqrt{2}\sqrt{N(N-1/4)}B}{R\sqrt{p}} \approx
\frac{\sqrt{2}N B}{R \sqrt{p}}
\label{equ:throughput}
\end{equation}

\begin{figure}[htb]
\begin{center}
\epsfig{file=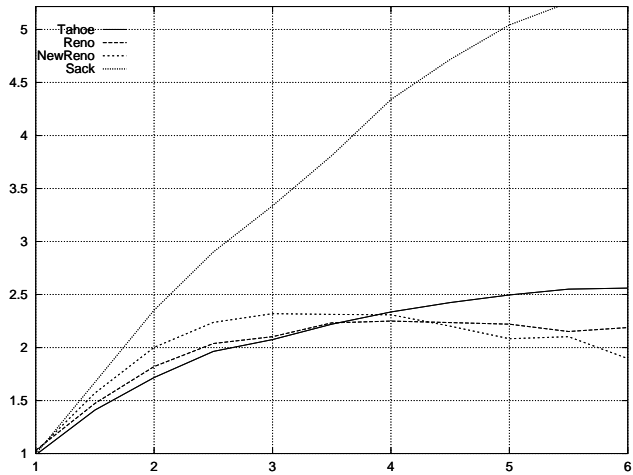,width=9cm}
\caption{Gain in throughput as a function of $N$}
\label{fig:multcp}
\end{center}
\end{figure}

Although the theoretical result looks good, practical
results from simulation are more interesting. In figure
\ref{fig:multcp} we have plotted the relative throughput of one MulTCP
connection against the throughput of a single connection. Both
connections share a bottleneck with 20 other TCP flows. The exact
setup of the simulation is given in Figure \ref{fig:network}. We have
applied the MulTCP extensions to four types of TCP, TCP Tahoe, TCP
Reno, New Reno and TCP Sack.

\begin{figure}
\begin{center}
\epsfig{file=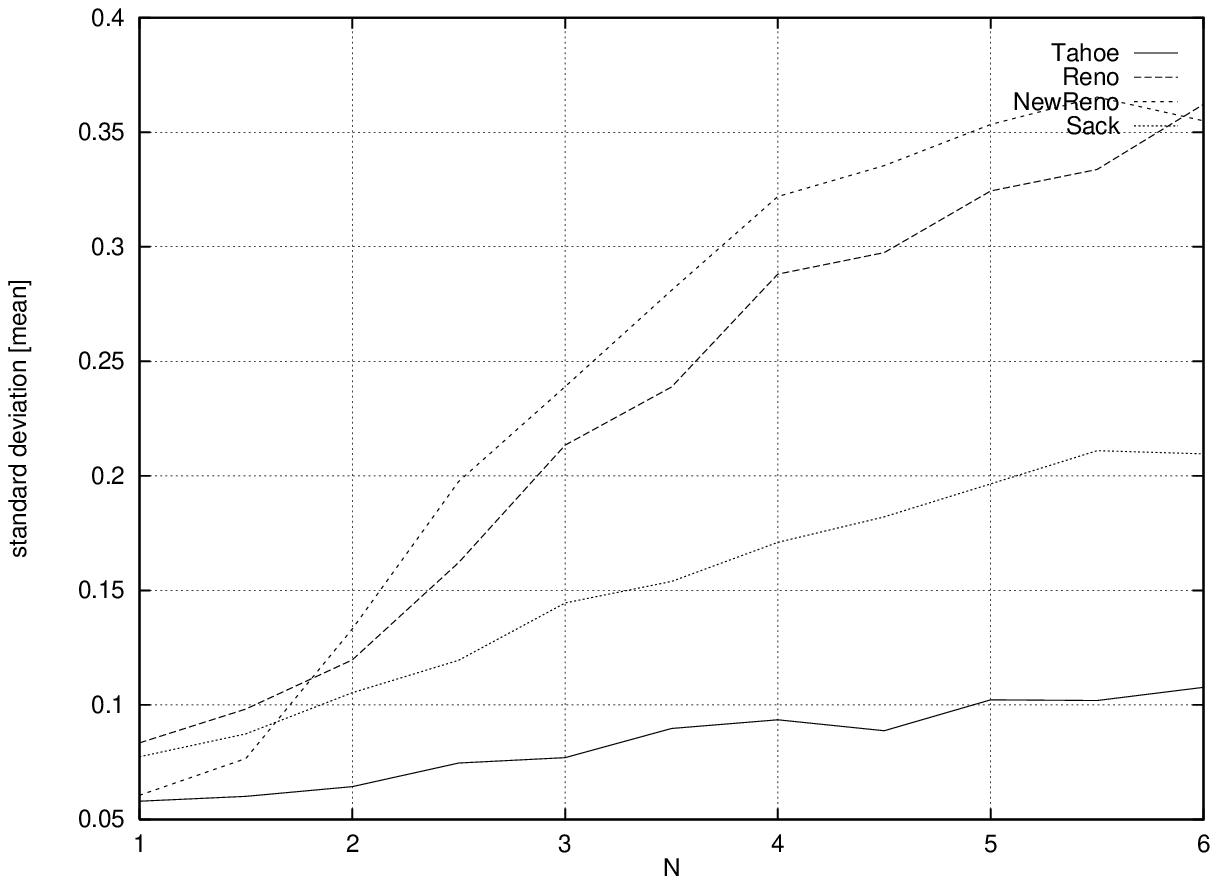,width=9cm}
\caption{Fairness among MulTCPs as a function of $N$}
\label{fig:fairness}
\end{center}
\end{figure}

For $N$ between one and two, a MulTCP flow gets about $N$ times the
throughput of one TCP flow. Except for TCP Sack, however, the
throughput does not go above $2.5$ for any larger $N$. This is due to
the fact that in our simulations all TCP flows experience timeouts now
and then. As we explained above, $N$ TCPs suffer less from timeouts
than one MulTCP. Thanks to its selective acknowledge mechanism, TCP
Sack can avoid most of the timeouts due to multiple errors. This is
why TCP Sack can increase its rate proportionally to $N$ up to a
factor of 10. However, there is a limit to this proportional increase
of the rate. There is only a fixed amount of information TCP Sack can
send in a selective acknowledgement. When $N$ is too high, MulTCP gets
too aggressive and and selective acknowledgement can not cope with
the multiple losses occurring. 

In an additional simulation we have investigated the effect of more
aggressive TCPs on fairness. The network we study is the same as for
the first simulation. This time however all the 22 connections have
the same parameter $N$. They should thus all get the same throughput
except for the bias due to different RTTs. For different values of $N$ we
have measured the throughput of each connection and multiplied each by
its RTT to normalise it. We then calculate the standard deviation of
the normalised throughput and express it in proportion of its mean
rate. The result are shown in Figure \ref{fig:fairness}. We see that
for $N$ small, the standard deviation is between $5$ and $10$ percent
of the mean rate. As $N$ increases, the standard deviation of the
throughput increases meaning that network is getting less and less
fair. 
This is probably due to the fact that more agressive TCPs are more
likely to generate bursts of losses which makes it more difficult for
the congestion window to stay close to its average size. Burst of
losses may also exarcerbate imperfections of the congestion control
mechanism.

\section{Billing and Policing} 
\label{sec:billing}

Billing is the set of procedures which are necessary for the network
provider to know how much to charge from a user. Policing on the other
hand allows the service provider to verify that the user really only
uses what he is paying for. The theory in \cite{kel97} calls for
billing of connections by duration and weight ($N$). In the case of
Web caches with receive buffer limited flows, billing can be done as a
byproduct of the receive buffer allocation, with hierarchical caches
allowing for aggregated billing. Policing is not an issue as the
device providing different quality of service is owned by the service
provider.

\begin{figure}
\begin{center}
\epsfig{file=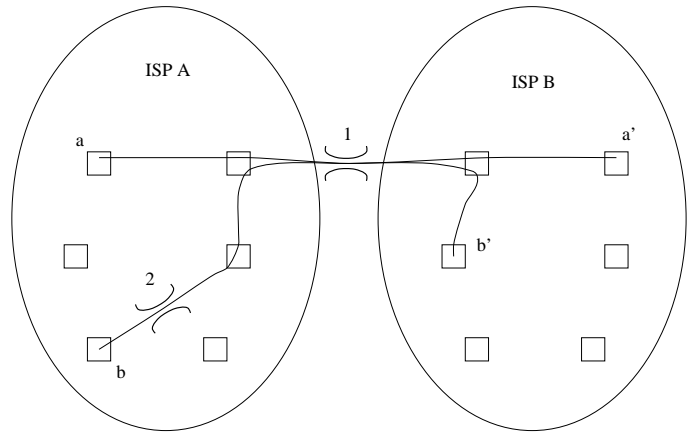,width=9cm}
\caption{Throughput and bottlenecks in a network. }
\label{fig:bottleneck}
\end{center}
\end{figure}

If proportional fairness is achieved through use of MulTCP, billing
and policing are more difficult. For example, measuring the rates at a
bottleneck does not give enough information to know how aggressive a
flow is. Indeed, as we see in Figure \ref{fig:bottleneck}, the fact
that a flow only uses a small portion of a bottleneck (1) can be due
to the fact that it has to cross a further bottleneck (2). To know
exactly what multiplier is being used on a TCP connection one either
needs information about all bottlenecks in the network or one can
analyse a trace of the flow. This allows to observe the number of
packets sent per acknowledgement during slow start and the variation
of the transmit window in presence of loss. For an example of a tool
doing this for standard TCPs we refer to \cite{pax97}. The task of
analysing all TCP connections is too complex and therefore we have to
use aggregation wherever possible.

We propose the following tentative method for billing and policing:

\begin{itemize}
\item{Policing:} Policing is done at random times on random
flows. MulTCP flows must declare the $N$ they are using by exchanging
a TCP options describing $N$ at connection setup. From a trace of a
connection one can verify that the flow did not behave more
agressively than it declared to be. By monitoring connection setups
one can deduct the average $N$ used by a user during a period of time.

\item{ISP-user billing:} At regular time interval the user declares
the average sum of $N$ of the connections run during the interval. The price is
calculated by multiplying the average sum of $N$ by the duration of
the interval. 

\item{ISP-ISP billing:}  At an exchange point, one ISP charges
another ISP for the total $N$ used by the traffic being
accepted. Indeed, the higher the sum of $N$ of the incoming
connections, the more resources will be tied up by that traffic. Again
this can be aggregated as the duration of a sampling period times the
average of the sum of $N$ of the incoming connections.
\end{itemize}





Having discussed the details of billing and policing we can now
propose an architecture for differentiated services in the Internet
using proportional fairness. 

\subsection{Weighted Proportional Fairness in the Internet, \\ a straw-man architecture}

This section describes how weighted proportional fairness could be
introduced into the current Internet using MulTCP, taking into acount
that it would have to coexist with standard best-effort service.

We limit the usage of MulTCP to users which generate a relatively high
amount of traffic, for example all users which have an access link to
the Internet with a throughput higher than a given value \footnote{
Note that HTTP and FTP make the major part of Internet traffic
            and that a large part of that traffic probably comes from
large servers}. Typical users are commercial Web servers. The
participating users set a bit in the Type Of Service (TOS) field of
the IP packets to indicate that they are part of the proportional
fairness scheme. When TCP connections are established with an $N$
larger than 1, the value of $N$ is transmitted as a TCP option in the
SYN message. The gateways in the network are provided with two queues,
one for best effort and one for Proportional Fair (PF) service. A
minimum amount of the network capacity is reserved for traffic of the
standard best effort service.
The users are requested to declare the average sum of all
multipliers per destination ISP at periodic intervals. This
information is sufficient for billing. The network provider charges
the network usage as the product of the sampling period
and the average sum of $N$ used in that period. If the user is a
Web server it can in turn charge the clients that downloaded content
or charge the advertisers that put advertisement in the content. For
policing purposes the ISP analyses the traces of random TCP
connections. If a connection is more aggressive than is indicated by
the $N$ in the TCP option, the user is penalised for not adhering to
the rules. For random time intervals the ISP calculates the average sum
of $N$ from the TCP options of the connections from one user to one
ISP. If this average doesn't match the average declared by the user,
the user is again penalised.
\footnote{In some scenarios, the unfairness users with larger RTTs experience
may not be the correct incentive. This is easily factored into
pricing directly from equation (1), but would need authenticated (policed) 
measurement. It may
be possible to estimate the RTT at a bottleneck but it is not trivial.
One can measure the RTT from the bottleneck to the server, and to the
client separately, by looking at the delay between transmission of
packet with a given sequence number and its acknowledgement;
but this would be for separate sequence number samples in each direction:
assuming uncorrelated delay distributions, one could combine these to form a reasonable
estimate for comparison of one flow's RTT with another.}

\section{Conclusions}

Weighted proportional fairness provides selective quality of service without
the need for connection acceptance control, reservations or multiple
queues in gateways. Moreover, as the network makes no explicit
promises to the user (other than who pays more gets more\footnote{
at least in the range where MulTCP really acts like $N$ TCPs}
) there is no
need for over provisioning. The total capacity of the network is always
available to its users and the price per bandwidth depends of the
instantaneous demand.

\paragraph{}
We have seen that the management of the receive buffers is one way to
implement weighted proportional fairness when all the flows share a bottleneck
and are terminated at the same host. This can be the case for example
in a system of Web cache servers. Weighted proportional fairness can also be
achieved by modifying TCP's congestion control algorithm. In that case
the range of the weight factor seems to be limited when TCPs 
don't use advanced techniques like selective acknowledgement to avoid
timeouts due to bursts of errors. The advantage of using the
congestion control algorithm as a means to achieve weighted proportional
fairness is that it can be done in a completely distributed manner and
independently of where the bottlenecks are located. 
\paragraph{}
In the absence of a policing and pricing scheme, we may see competition between
different TCP implementations. It is clear from this and other work
that a 'tweaked' Sack-TCP can be more aggressive than prior TCPs, and 
still be stable. This
implies that we need to police SACK users anyway, to be fair to older
TCPs. Of course, there should be some incentive for people to migrate
implementations to more effective protocol mechanisms too, but not so
that they also increase their network share under cover of the move,
above that achieved by efficiency savings natural to the protocol!
\paragraph{}
Finally, we conjecture that while distributed control scales well, it
leads to non-scalable policing at distributed bottlenecks. However,
the converse may be true for max-min fairness schemes, where policing scales 
(i.e. aggregates) but control schemes do not scale so well
(i.e. require distributed ``n-squared'' agreement during signalling).

\subsection{Acknowledgements}
The authors would like to thank Frank Kelly for many fruitful discussions
and comments on the draft of this paper. We would also like to thank the 
anonymous reviewers for points of clarification.

\bibliographystyle{plain}
\bibliography{papers}

\newpage

\section*{Annex A: Min-max Fairness, Proportional Fairness and
Weighted Proportional Fairness}

Alternative notions of fairness arise in several disciplines, from
political philosophy \cite{raw71} to communication engineering
\cite{ber87}. In this Annex we formally define max-min fairness,
proportional fairness, and weighted proportional fairness.

Let $S$ be a set of connections and the vector $x = (x_s, s \in S)$
the rate of each connection. Define $x$ to be feasible in a network
$N$ if all rates are non-negative and if the sum of rates on each link
of the network does not exceed the links capacity.

\paragraph{max-min fairness:} a vector of rates $x$ is max-min fair if
for any other feasible vector $y$, there exists $r$ such that $y_r > x_r$
implies that there exists $s$ such that $y_s < x_s < x_r$.


\paragraph{}For a discussion of max-min fairness in a variety of contexts the
reader is referred to \cite{raw71} and \cite{ber87}.

\paragraph{proportional fairness:} a vector of rates $x$ is
proportionally fair if it is feasible and if for any other feasible
vector $y$, the aggregate of proportional changes is zero or negative:
\[\sum_{s\in S}\frac{y_s - x_s}{x_s} \leq  0\]

If x is proportionally fair, then it is the Nash bargaining solution,
satisfying certain axioms of fairness \cite{gar95}.

\paragraph{} Let $w = (w_s, s \in S)$ be a vector of weights, or charges.

\paragraph{weighted proportional fairness:} a vector of rates $x$ is
proportionally fair per unit charge if it is feasible and if for any
other feasible vector $y$,
\[\sum_{s\in S}w_s\frac{y_s - x_s}{x_s} \leq  0\]
 
For a discussion of weighted proportional fairness and its 
relation to utility maximisation, see \cite{kel97a,kel97}.

\section*{Annex B: Steady state throughput of MulTCP}

To approximate the steady state throughput of a MulTCP flow we use the
same approach as in \cite{flo97}. We assume that the congestion window $W$ 
varies in a saw-tooth shape. When a loss occurs it is reduced to $W
\frac{N-1/2}{N}$. It then grows by $N$ per round trip time until it
reaches its original size and a new loss occurs. The amount of data
transmitted during one cycle is thus: 
\[ S = W\frac{N-1/2}{N} + \left(W\frac{N-1/2}{N} + N \right) + ... + W
\approx \frac{W^2}{2N^2}\frac{N-1/4}{N} \]

which is the inverse of the loss rate as one packet is lost per cycle:
\begin{equation} 
 p = \frac{2N^2}{W^2}\frac{N}{N-1/4}
\label{eq:p}
\end{equation}

The throughput is equal to the average congestion window size times
the size of a packet $B$ and divided by the round trip time $R$:
\[ T(N) = \frac{\overline{W} B}{R} \]

Now $\overline{W}$ is equal to $W\frac{N-1/4}{N}$ and $W$ can be
substituted from Equation \ref{eq:p} yielding:

\begin{equation}
T =  \frac{\sqrt{2}\sqrt{N(N-1/4)}B}{R\sqrt{p}} \approx
\frac{\sqrt{2}N B}{R \sqrt{p}}
\end{equation}

For $N = 1$ we get the same result as in \cite{flo97}:
\[ T_1 = \frac{\sqrt{3/2} \: B}{R \: \sqrt{p}} \]

We thus have:
\begin{eqnarray*}
  T & = & \frac{2}{\sqrt{3}}\sqrt{N(N-1/4)} \: T_1 \\
    & \approx & N \: T_1 \:{\rm \ for\  }\: 1 \leq N \leq 10
\end{eqnarray*}

\end{document}